\documentclass[pra,aps,twocolumn,amsmath,amssymb,showkeys]{revtex4}
\usepackage{graphicx}
\newcommand{\hh}{{\mathcal{H}}}
\newcommand{\cli}{{\mathcal{I}}}

\newcommand{\mtd}{{\mathsf{D}}}
\newcommand{\mtm}{{\mathsf{L}}}
\newcommand{\mse}{{\mathsf{E}}}
\newcommand{\msg}{{\mathsf{G}}}
\newcommand{\msh}{{\mathsf{H}}}
\newcommand{\msj}{{\mathsf{J}}}
\newcommand{\msk}{{\mathsf{K}}}
\newcommand{\mtx}{{\mathsf{X}}}
\newcommand{\mty}{{\mathsf{Y}}}
\newcommand{\mtz}{{\mathsf{Z}}}
\newcommand{\bro}{\boldsymbol{\rho}}
\newcommand{\vbro}{\boldsymbol{\varrho}}
\newcommand{\bmg}{\boldsymbol{\omega}}
\newcommand{\bsg}{\boldsymbol{\sigma}}
\newcommand{\bdl}{\boldsymbol{\delta}}
\newcommand{\Upn}{\Upsilon}
\newcommand{\vcr}{{\mathbf{r}}}
\newcommand{\pen}{\openone}
\newcommand{\tr}{\mathrm{tr}}
\newcommand{\dig}{\mathrm{diag}}
\newcommand{\rk}{\mathrm{rank}}

\newcommand{\suc}{\mathrm{suc}}

\newcommand{\ron}{{\mathrm{ran}}}
\newcommand{\clm}{{\mathcal{M}}}
\newcommand{\cpt}{{\mathtt{C}}}
\newcommand{\iu}{{\mathtt{i}}}

\begin{document}
\clearpage
\preprint{}

\title{On degradation of Grover's search under collective phase flips in queries to the oracle}

\author{Alexey E. Rastegin}
\affiliation{Department of Theoretical Physics, Irkutsk State University, Russia,
e-mail: alexrastegin@mail.ru \\}

\begin{abstract}
We address the case, when querying to the oracle in Grover's
algorithm is exposed to noise including phase distortions. The
oracle-box wires can be altered by an opposite party that tries to
prevent correct receiving data from the oracle. This situation
reflects an experienced truth that any access to prophetic
knowledge cannot be common and direct. To study the problem, we
introduce a simple model of collective phase distortions on the
base of phase damping channel. In the model used, the success
probability is not altered via the oracle-box wires {\it per se}.
Phase distortions of the considered type can hardly be detected
via any one-time query to the oracle. However, the success
probability is significantly changed, when such errors are
introduced as an intermediate step into the Grover iteration. We
investigate the success probability with respect to variations the
parameter that characterizes the amount of phase errors. It turns
out that the success probability is decreased significantly even
if the error amount is not very high. Moreover, this probability
quickly reduces to the value of one half, which corresponds to the
completely mixed state. We also study trade-off relations between
quantum coherence and the success probability in the presence of
noise of the considered type.
\end{abstract}

\keywords{Grover's algorithm, phase noise, relative entropy of coherence}

\maketitle

\pagenumbering{arabic}
\setcounter{page}{1}

\section{Introduction}\label{sec1}

In last decades, quantum effects have found a new field of
applications for information processing \cite{mardel02}. Famous
Shor's discovery \cite{shor97} had lead to numerous quantum
algorithms for algebraic problems
\cite{hama2006,hallg07,vandam2010}. Grover's search algorithm
\cite{grover97,grover97a,grover98} is another fundamental result.
Inspired amplitude amplification technique is widely used as one
of primary tools in building quantum algorithms \cite{patel2016}.
The algorithms of Shor and Grover may be related more closely than
it seems initially \cite{loka2007}. Studying of quantum algorithms
is a part of recent efforts to realize emerging technologies in
quantum information processing. The Grover algorithm is optimal
for search by means of queries to the oracle
\cite{bbbv97,zalka99}. The user invoke the oracle to process any
item, whereas the database itself is not represented explicitly.
The original formulation has been modified with particular blocks
of more general kind. In addition, amplification process may start
with an arbitrary initial distribution of the amplitudes.
Generalized versions of Grover's algorithm were thoroughly
analyzed in \cite{biham99,mardel00,biham2000,biham2002}.

Algorithms of amplitude amplification are related to the case,
when users employ an access to the so-called oracle. The oracle
denotes some black box, which is able to calculate values of the
desired Boolean function. Users query the box by putting concrete
values of the argument. In reality, an access to the oracle may be
impeded and unreliable. Due to a wide applicability of amplitude
amplification technique, this problem deserves detailed
investigations. The oracle-box wires are inevitably exposed to
noise, even if its amount is low. In addition, the wires could be
affected due to activity of an opposite party. There are many
possible scenarios, in which the above questions could be
examined. In this work, we address one of such scenarios. Despite
of its simplicity, we disclose some unexpected corollaries of
collective phase flips in the oracle-box wires. In the model
considered, queries to the oracle are exposed to phase flips
described similarly to the phase damping of a qubit. We also
discuss the relative entropy of coherence from the viewpoint of
its trade-off with the success probability.

The paper is organized as follows. The preliminary material is
given in Sect. \ref{sec2}. In Sect. \ref{sec3}, we introduce the
model of collective flips that occurs in the oracle-box wires. The
used model leads to the recursion equation in terms of the
effective Bloch vector. From the viewpoint of noise amount, two
certain cases should be distinguished.  In Sect. \ref{sec5}, we
examine changes of the success probability after repeated Grover's
iterations under collective phase flips. It turns out that the
Grover search algorithm is very sensitive to distortions of the
considered type. Using the relative entropy of coherence, we
further study trade-off relations with the success probability. In
Sect. \ref{sec6}, we conclude the paper with a summary of results.
Appendix \ref{app1} is devoted to solutions of the derived
recursion equation.

\section{Preliminaries}\label{sec2}

In this section, the required material will be recalled. We begin
with a lot of linear algebra. Some aspects of the Grover search
algorithm, quantum noise and coherence measures will also be
presented. For a rectangular $m\times{n}$ matrix $\mtz$, its
singular values $s_{j}(\mtz)$ are defined as square root of the
eigenvalues of positive semidefinite matrix $\mtz^{\dagger}\mtz$
\cite{watrous1}. The number of non-zero singular values is equal
to $\rk(\mtz)\leq\min\{m,n\}$. The matrices $\mtz^{\dagger}\mtz$
and $\mtz\mtz^{\dagger}$ have the same non-zero eigenvalues. For
$q\in[1;\infty]$, the Schatten $q$-norm is defined as
\cite{watrous1}
\begin{equation}
\|\mtz\|_{q}:=\left(
\sum\nolimits_{j=1}^{\rk(\mtz)} s_{j}(\mtz)^{q}
\right)^{\!1/q}
 . \label{shmq}
\end{equation}
In the following, we will use the Frobenius norm
$\|\mtz\|_{2}:=\sqrt{\tr(\mtz^{\dagger}\mtz)}$ and the spectral
norm $\|\mtz\|_{\infty}:=\max{s}_{j}(\mtz)$. One of important
properties of the Schatten norms is expressed by the inequality
(see, e.g., formula (1.175) in \cite{watrous1})
\begin{equation}
\|\mtx\mty\mtz\|_{q}\leq\|\mtx\|_{\infty}\,\|\mty\|_{q}\,\|\mtz\|_{\infty}
\ . \label{w1nq}
\end{equation}

Let us recall the original formulation of Grover's search
algorithm. The search space contains $N=2^{n}$ items denoted by
binary $n$-string $x=(x_{1}\cdots{x}_{n})$ with $x_{j}\in\{0,1\}$
so that $x\in\{0,1,\ldots,N-1\}$. The problem is to find one of
marked items that form some set $\clm$. By $\clm^{\cpt}$, we mean
the complement of this set. Without loss of
generality, we can assume $1\leq|\clm|\leq{N}/2$.

Checking items, the algorithm appeals to the so-called ``oracle''.
For the each given $x$, the oracle returns the value of Boolean
function $x\mapsto{F}(x)$ such that $F(x)=1$ for $x\in\clm$ and
$F(x)=0$ for $x\in\clm^{\cpt}$. Initially, the algorithm
initializes the $n$-qubit register to $|0\rangle$. Applying the
Hadamard transform, one gets the distribution with equal
amplitudes, namely
\begin{equation}
\msh|0\rangle=\frac{1}{\sqrt{N}}\sum_{x=0}^{N-1} |x\rangle
\, . \label{uniam}
\end{equation}
Such superpositions are used to realize the quantum parallelism
\cite{deutsch85}. Furthermore, we repeat the Grover iteration
involving two steps. The first step with querying the oracle can
be represented by the rotation operator
\begin{equation}
\msj=\sum_{x=0}^{N-1} (-1)^{F(x)}\,|x\rangle\langle{x}|
\, . \label{jfpi}
\end{equation}
By (\ref{jfpi}), amplitudes of marked states are all multiplied by
the phase factor $\exp(\iu\pi)=-1$. This step has been generalized
to other values of the phase \cite{biham99,biham2000}. Since we
shall focus on an influence of noise, such generalizations are not
considered in the following. The second step of the Grover
iteration realizes the inversion about mean \cite{nielsen}. It is
described by the operator
\begin{equation}
\msk=2\msh|0\rangle\langle0|\msh-\pen_{N}
\, , \label{roj0}
\end{equation}
where $\pen_{N}$ is the identity operator of the corresponding
size. We can sometimes replace (\ref{roj0}) with a more general
block. Such algorithms were analyzed in \cite{biham99,biham2000}.
Thus, the standard Grover iteration is written as
\begin{equation}
\msg=\msk\msj
\, . \label{msg0}
\end{equation}
In the standard formulation, the initial distribution of
amplitudes is taken in the form (\ref{uniam}). Then
the evolution of amplitudes can be described within the
two-dimensional picture. Let us define the normalized
superpositions of unmarked and marked states,
\begin{align}
|w\rangle&:=\frac{1}{\sqrt{N-M}}\sum_{x\in\clm^{\cpt}} |x\rangle
\, . \label{sunm}\\
|m\rangle&:=\frac{1}{\sqrt{M}}\sum_{x\in\clm} |x\rangle
\, , \label{smar}
\end{align}
We will also use the parameter $\theta\in(0;\pi/2)$ such that
$\cos\theta=1-2M/N$, whence
\begin{equation}
\sin^{2}\theta/2=\frac{M}{N}
\ , \qquad
\cos^{2}\theta/2=1-\frac{M}{N}
\ . \label{sicos0}
\end{equation}
Writing operators as matrices in the basis
$\bigl\{|w\rangle,|m\rangle\bigr\}$, we have
$\msj=\dig(+1,-1)=\bsg_{z}$ and
\begin{equation}
\msk=
\begin{pmatrix}
    \cos\theta & \sin\theta \\
    \sin\theta & \,-\cos\theta
\end{pmatrix}
 . \label{kmat}
\end{equation}
Thus, the operator (\ref{msg0}) is simply represented as
\cite{nielsen}
\begin{equation}
\msg=
\begin{pmatrix}
    \cos\theta & \,-\sin\theta \\
    \sin\theta & \cos\theta
\end{pmatrix}
 . \label{grot0}
\end{equation}
Thus, each Grover iteration rotates the register state by $\theta$
towards the superposition $|m\rangle$. We wish to address the case
when querying to the oracle is exposed to noise. To obtain
explicit results, we have to restrict a consideration on
sufficiently simple models of errors.

Entanglement is one of key resources in quantum computations. Due
to findings of the papers \cite{bpati2002,jozsa03}, quantum
speed-up without entanglement is hardly possible. To analyze the
nature of quantum algorithms, we should think about correlations
in the context of few prescribed bases. This question is related
to changes of quantum coherence of the register during
computational processes. The framework for studies of coherence as
purely quantum feature was developed in recent years
\cite{bcp14,plenio16}. Concerning amplitude amplification,
trade-off relations between quantum coherence and the success
probability is one of important questions. We shall address this
question in the case, when queries to the oracle are exposed to
phase flips of the considered type. To do so, the relative entropy
of coherence will be utilized.

In general, various approaches to measure quantum correlations are
discussed in \cite{abc16,plenio16,hufan16,fan2017}. The authors of
\cite{bcp14} developed a list of axioms that should be satisfied
by any proper quantifier of coherence. As a rule, each candidate
to quantify the amount of coherence is associated with some
distinguishability measure. Let us take the set $\cli$ of all
diagonal density matrices, namely
\begin{equation}
\bdl=\sum_{x=0}^{N-1} b(x)\,|x\rangle\langle{x}|
\, , \qquad
\sum_{x=0}^{N-1} b(x)=1
\, . \label{incd}
\end{equation}
We asks how far the given state is from those states that are
completely incoherent in the computational basis. Using the
quantum relative entropy as a measure of distinguishability leads
to the relative entropy of coherence. The quantum relative entropy
of $\bro$ with respect to $\bmg$ is defined as
\cite{watrous1,vedral02}
\begin{equation}
D_{1}(\bro||\bmg):=
\begin{cases}
\tr(\bro\ln\bro\,-\bro\ln\bmg) \,,
& \text{if $\ron(\bro)\subseteq\ron(\bmg)$} \, , \\
+\infty\, , & \text{otherwise} \, .
\end{cases}
\label{relan}
\end{equation}
By $\ron(\bro)$, we mean here the range of $\bro$. On the base of
(\ref{relan}), the corresponding coherence measure is introduced as
\cite{bcp14}
\begin{equation}
C_{1}(\bro):=
\underset{\bdl\in\cli}{\min}\,D_{1}(\bro||\bdl)
\, . \label{c1df}
\end{equation}
The minimization leads to the expression \cite{bcp14}
\begin{equation}
C_{1}(\bro)=S_{1}(\bro_{\dig})-S_{1}(\bro)
\, , \label{c1for}
\end{equation}
where $S_{1}(\bro)=-\,\tr(\bro\ln\bro)$ is the von Neumann entropy
of $\bro$, and
\begin{equation}
\bro_{\dig}:=\sum_{x=0}^{N-1} p(x)\,|x\rangle\langle{x}|
\, ,
\qquad
p(x)=\langle{x}|\bro|x\rangle
\, . \nonumber
\end{equation}
The entropy $S(\bro_{\dig})$ is equal to the Shannon entropy
calculated with the probabilities $p(x)$. Basic properties of
(\ref{c1df}) are considered in \cite{bcp14,plenio16}. Generalized
entropic functions have found use in quantum information theory.
It is for this reason that we designate the above quantities by
the subscript $1$. Coherence quantifiers induced by quantum
divergences of the Tsallis type were addressed in
\cite{rastpra16}. It was shown that such quantifiers do not allow
a simple form similar to (\ref{c1for}). Coherence monotones based
on R\'{e}nyi divergences were considered in
\cite{chitam2016,shao16,skwgb16}. Other candidates to quantify the
amount of coherence were examined in \cite{shao15,rpl15}. The
geometric coherence is an interesting quantifier of different
character \cite{plenio16}. Two-sided estimates on the geometric
coherence were obtained in \cite{geomes17}.

Complementarity relations for quantum coherence can be formulated
in several ways \cite{hall15,pati16,pzflf16,baietal6,rastcomu}.
Duality relations between the coherence and path information were
examined in \cite{bera15,bagan16,qureshi17}. From the viewpoint of
quantum computations, the concept of quantum coherence was studied
in \cite{hillery16,hfan2016,apati2016}. In particular, the authors
of \cite{hfan2016} reported on coherence depletion in the original
Grover algorithm. In the paper \cite{rastgro17}, we have studied
relations between coherence and the success probability in
generalized amplitude amplification. Some of these results will be
used in studies of Grover's search in the presence of phase flips.

\section{Collective flips introduced by phase damping}\label{sec3}

There are infinitely many scenarios of an interaction with
environment. We will consider a model with phase damping. Such
processes describe the loss of quantum information without loss of
energy \cite{nielsen}. It gives a ground for understanding
physical effects in quantum systems similar to the Schr\"{o}dinger
cat-atom system. Let us focus on those density matrices that are
effectively two-dimensional with respect to the basis
$\bigl\{|w\rangle,|m\rangle\bigr\}$. In other words, they can be
represented via usual Bloch vector $\vcr=(r_{x},r_{y},r_{z})$,
namely
\begin{equation}
\bro=\frac{1}{2}
\begin{pmatrix}
    1+r_{z} & r_{x}-\iu{r}_{y} \\
    r_{x}+\iu{r}_{y} & 1-r_{z}
\end{pmatrix}
 . \label{dmatr}
\end{equation}
With positive parameter $\eta\leq1$, we introduce the following
Kraus operators,
\begin{equation}
\mse_{0}:=
\begin{pmatrix}
    1 & 0 \\
    0 & \sqrt{\eta}
\end{pmatrix}
 , \qquad
\mse_{1}:=
\begin{pmatrix}
    0 & 0 \\
    0 & \sqrt{1-\eta}
\end{pmatrix}
 . \label{kraus01}
\end{equation}
These operators prescribe the action of phase damping
$\Phi_{\mse}$ on density matrices of the above type. It is well
known that this action reads as \cite{nielsen}
\begin{equation}
(r_{x},r_{y},r_{z})\overset{\Phi_{\mse}\,}\longmapsto
\bigl(\sqrt{\eta}\,r_{x},\sqrt{\eta}\,r_{y},r_{z}\bigr)
\, . \label{blov}
\end{equation}
Initializing gives the density matrix
$\bro(0)=\msh|0\rangle\langle0|\msh$ with the Bloch vector
$\vcr(0)=(\sin\theta,0,\cos\theta)$. After $t$ iterations, the
success probability is written as
\begin{equation}
P_{\suc}(t)=\langle{m}|\bro(t)|m\rangle=\frac{1-r_{z}(t)}{2}
\ . \label{psuct}
\end{equation}
It must be stressed that the channel $\Phi_{\mse}$ itself cannot
alter the probability of the success. We immediately see this fact
from (\ref{blov}). Nevertheless, the probability of the success
may be changed in the case, when this channels acts as an
intermediate point in amplitude amplification process.

We will investigate the following scheme. Each iteration
transforms density matrices of the register according to the
formula
\begin{equation}
\bro(t)\mapsto\bro(t+1)
=\Upn_{\msk}\circ\Phi_{\mse}\circ\Upn_{\msj}\circ\Phi_{\mse}\bigl(\bro(t)\bigr)
\, , \label{iteralt}
\end{equation}
where the two unitary channels are written as
\begin{equation}
\Upn_{\msj}(\vbro)=\msj\vbro\msj^{\dagger}
\, , \qquad
\Upn_{\msk}(\vbro)=\msk\vbro\,\msk^{\dagger}
\, . \label{unikj}
\end{equation}
For $\eta=1$, the map $\Phi_{\mse}$ reads as identical, so that
the right-hand side of (\ref{iteralt}) takes the form
\begin{equation}
\bro(t+1)=\Upn_{\msk}\circ\Upn_{\msj}\bigl(\bro(t)\bigr)
=\msg\bro(t)\msg^{\dagger}
\, . \label{iterc}
\end{equation}
If the initialized state is pure, then the register will remain to
be in pure states under the action of the map (\ref{iterc}). It is
not the case for the altered map (\ref{iteralt}).

The above model deals with collective phase distortions that are
simply expressed in the computational basis. It allows us to
formulate results in a closed analytic form. Under some
circumstances, the considered picture concerns the case when phase
flips occur in a single qubit solely. The phase flip channel has
Kraus operators $\sqrt{\alpha}\,\pen_{2}$ and
$\sqrt{1-\alpha}\,\bsg_{z}$ \cite{nielsen}. This channel is actually
equivalent to the phase damping channel, whenever
\begin{equation}
2\alpha=1+\sqrt{\eta}
\, . \label{aleta}
\end{equation}
If only one of qubits is affected, then we can split the total
Hilbert space into two subspaces $\hh_{0}$ and $\hh_{1}$. These
subspaces are spanned by canonical states that have in their
binary notation either $0$ or $1$ in the position of the noised
qubit, respectively. Then any density matrix $\bro$ can be written
as a block $2\times2$-matrix $[[\bro_{ij}]]$. Due to phase flips
in the noised qubit, we have
$\bro_{ij}\mapsto\sqrt{\eta}\,\bro_{ij}$ for $i\neq{j}$. The
diagonal submatrices $\bro_{00}$ and $\bro_{11}$ are not changed.
Suppose that errors occur in each query to the oracle. In general,
a complete analysis is complicated \cite{reitzner17}. However, the
situation can sometimes be reduced to our model. When one of
subspaces consists of only marked states, we really deal with
the channel $\Phi_{E}$ due to (\ref{blov}). Our scenario does not
mean that phase distortions act in all qubits simultaneously.
Dephasing noise may be treated as a result of the existence of
``damaged'' vertices in quantum-walk search \cite{reitzner17}. Due
to an interferometric picture of quantum walks \cite{hbf03}, one
can therefore give an example of phase errors in realistic
systems.

Interpreting (\ref{iteralt}) as a matrix relation, we will solve
it via the diagonalization. This approach is somehow similar to
the treatment of \cite{biham2000}. On the other hand, we deal with
the recursion equation for components of the actual Bloch vector.
The authors of \cite{biham2000} used the recursion equation for
components of the wave function. In our case, only two of
components of the Bloch vector are non-zero. So, we will further
treat $\vcr(t)$ as a column with two entries, namely $r_{x}(t)$
and $r_{z}(t)$. In terms of the Bloch vector components, the
action of operation $\Phi_{\mse}$ is represented by the matrix
$\dig(\sqrt{\eta},+1)$. We also have
\begin{align}
2\msj\bro\msj^{\dagger}&=\bsg_{z}(\pen_{2}+r_{x}\bsg_{x}+r_{z}\bsg_{z})\bsg_{z}
\nonumber\\
&=\pen_{2}-r_{x}\bsg_{x}+r_{z}\bsg_{z}
\, , \label{bsxz}
\end{align}
whence the operation
$\Upn_{\msj}$ acts on the Bloch vectors as the matrix
$\dig(-1,+1)$. Using (\ref{kmat}), we obtain
$\msk\pen_{2}\,\msk^{\dagger}=\pen_{2}$,
\begin{align}
\msk\bsg_{x}\msk^{\dagger}&=-\cos2\theta\,\bsg_{x}+\sin2\theta\,\bsg_{z}
\, , \nonumber\\
\msk\bsg_{z}\msk^{\dagger}&=\sin2\theta\,\bsg_{x}+\cos2\theta\,\bsg_{z}
\, , \nonumber
\end{align}
so that $r_{x}\mapsto-\cos2\theta\,r_{x}+\sin2\theta\,r_{z}$ and
$r_{z}\mapsto\sin2\theta\,r_{x}+\cos2\theta\,r_{z}$ due to the
operation $\Upn_{\msk}$. Hence, this operation acts on the Bloch
vectors as the matrix
\begin{equation}
\begin{pmatrix}
    -\cos2\theta & \sin2\theta \\
    \sin2\theta & \cos2\theta
\end{pmatrix}
 . \label{kvbk}
\end{equation}
Finally, the recursion equation is written in terms of the
effective Bloch vector as
\begin{equation}
\vcr(t+1)=\mtm\,\vcr(t)
\, , \label{itrr}
\end{equation}
where the matrix $\mtm$ reads as
\begin{equation}
\mtm=
\begin{pmatrix}
    \eta\,\cos2\theta & \sin2\theta \\
    -\,\eta\,\sin2\theta & \cos2\theta
\end{pmatrix}
 . \label{matra}
\end{equation}
Thus, we obtain the equation $\vcr(t)=\mtm^{t}\,\vcr(0)$ solved
explicitly in Appendix \ref{app1}. The singular values of $\mtm$
are equal to $1$ and $\eta$, whence $\|\mtm\|_{\infty}=1$.
Combining the latter with (\ref{w1nq}) implies that the $2$-norm
of the Bloch vector cannot increase, i.e.,
$\|\vcr(t)\|_{2}\leq\|\vcr(0)\|_{2}$.

It will be convenient to put three positive parameters, viz.
\begin{align}
A_{\pm}&:=\frac{1\pm\eta}{2}\,\cos2\theta
\, , \label{adef}\\
B&:=
\begin{cases}
\sqrt{\eta-A_{+}^{2}} \, ,
& \text{if } \eta>A_{+}^{2} \, , \\
0 \, , & \text{if } \eta=A_{+}^{2}\, , \\
\sqrt{A_{+}^{2}-\eta}\, ,
& \text{if } \eta<A_{+}^{2} \, .
\end{cases}
\label{bdef}
\end{align}
The characteristic equation is written as
\begin{equation}
\lambda^{2}-2A_{+}\lambda+\eta=0
\, . \label{chareq}
\end{equation}
We will further assume that $A_{+}^{2}-\eta\neq0$. Then the
eigenvalues $\lambda_{+}$ and $\lambda_{-}$ differ so that the
matrix $\mtm$ is certainly diagonalizable. The following two cases
should be mentioned: (i) $\eta>A_{+}^{2}\,$; (ii)
$\eta<A_{+}^{2}\,$. In Appendix \ref{app1}, we examine them
separately. In the first case, one has
\begin{align}
r_{z}(t)=\frac{\eta^{t/2}}{B}\>
&\Bigl(
-\,\eta\,\sin\varphi{t}\,\sin2\theta\,\sin\theta
\Bigr.
\nonumber\\
&\Bigl.
{}+(B\cos\varphi{t}+A_{-}\sin\varphi{t})\cos\theta
\Bigr)
\, , \label{arzt0}\\
P_{\suc}(t)=\frac{1}{2B}\>
&\Bigl(
B+\eta^{1+t/2}\sin\varphi{t}\,\sin2\theta\,\sin\theta
\Bigr.
\nonumber\\
&\Bigl.
{}-\eta^{t/2}(B\cos\varphi{t}+A_{-}\sin\varphi{t})\cos\theta
\Bigr)
\, . \label{psct0}
\end{align}
In the second case, one gets
\begin{align}
r_{z}(t)=\frac{\eta^{t/2}}{B}\>
&\Bigl(
-\,\eta\,\sinh\phi{t}\,\sin2\theta\,\sin\theta
\Bigr.
\nonumber\\
&\Bigl.
{}+(B\cosh\phi{t}+A_{-}\sinh\phi{t})\cos\theta
\Bigr)
\, , \label{arzt}\\
P_{\suc}(t)=\frac{1}{2B}\>
&\Bigl(
B+\eta^{1+t/2}\sinh\phi{t}\,\sin2\theta\,\sin\theta
\Bigr.
\nonumber\\
&\Bigl.
{}-\eta^{t/2}(B\cosh\phi{t}+A_{-}\sinh\phi{t})\cos\theta
\Bigr)
\, . \label{psct}
\end{align}

\section{On dynamics of the success probability and quantum coherence}\label{sec5}

In this section, we will use the solutions (\ref{psct0}) and
(\ref{psct}) for studying a vulnerability of Grover's search with
respect to phase flips in the oracle-box wires. In principle, they
may be inspired by an opposite party that tries to prevent correct
querying to the oracle. To input collective phase flips, the
opponent should be aware of details of the Boolean function
$x\mapsto{F}(x)$.

\begin{figure}
\includegraphics[width=7.8cm]{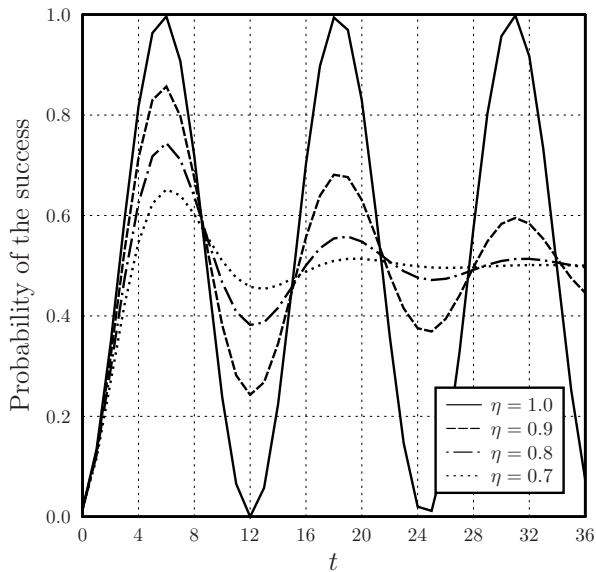}
\caption{\label{fig1} In the case (i), $P_{\suc}(t)$ is shown for
$N=64$, $M=1$, and the four values of $\eta$.}
\end{figure}

To study significance of collective phase flips, we visualize
$P_{\suc}(t)$ versus $t$ for several values of the parameter
$\eta$. We begin with the case (i), when $\eta>A_{+}^{2}\,$. As
was mentioned above, this case includes the standard situation
$\eta=1$. In Fig. \ref{fig1}, we take $N=64$, $M=1$, and show the
lines for the four values of $\eta$. Although $t$ takes integer
values, one draws lines as continuous for the sake of visibility.
Even if the amount of errors is low, values of the success
probability are reduced essentially. Without distortions, when
$\eta=1$, the dependence of $P_{\suc}(t)$ on $t$ is almost
periodic. Noticeable decreasing of $P_{\suc}(t)$ due to $\eta<1$
is observed even in the first cycle of amplitude amplification.
After several cycles, the curve $P_{\suc}(t)$ asymptotically
degenerates to the constant equal to $1/2$. The same result was
later reported for a particular case of decoupling noise in
\cite{reitzner17}.

\begin{figure}
\includegraphics[width=7.8cm]{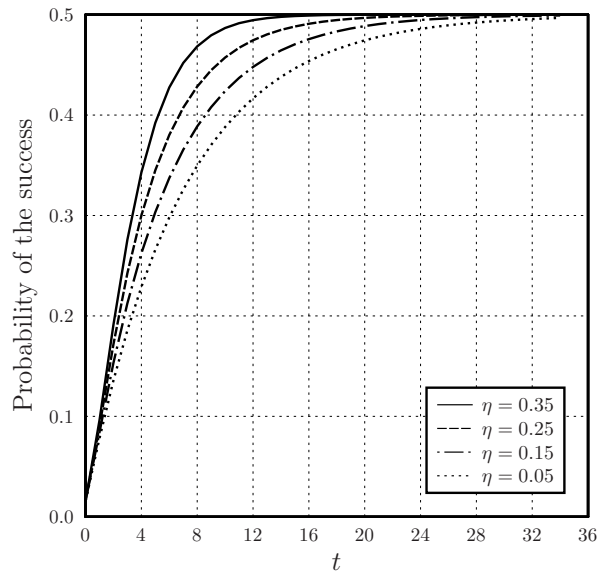}
\caption{\label{fig2} In the case (ii), $P_{\suc}(t)$ is shown for
$N=64$, $M=1$, and the four values of $\eta$.}
\end{figure}

For the completeness, we also show an example of $P_{\suc}(t)$ for
the case (ii), when $\eta<A_{+}^{2}\,$. In Fig. \ref{fig2}, we set
$N=64$, $M=1$, and show the lines for the four low values of
$\eta$. Here, the dependence $P_{\suc}(t)$ is of a completely
different character. Instead of decaying cycles with some peaks,
we observe that the lines smoothly and quickly saturate the
constant equal to $1/2$. This behavior cannot be treated as
amplitude amplification of any kind. When the oracle-box wires are
exposed to the channel $\Upn_{\msk}$ with such low values $\eta$,
legitimate users are rather able to detect this fact. So, we
further return to the case (i) and address trade-off relations
between quantum coherence and the success probability.

\begin{figure}
\includegraphics[width=7.8cm]{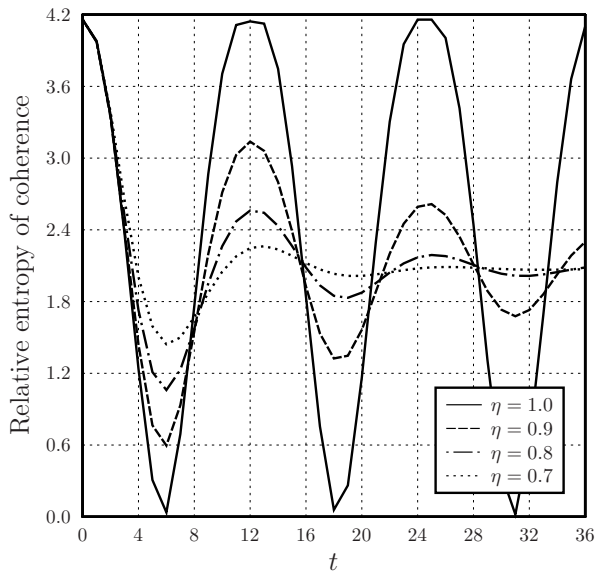}
\caption{\label{fig3} In the case (i), $C_{1}\bigl(\bro(t)\bigr)$
is shown for $N=64$, $M=1$, and the four values of $\eta$.}
\end{figure}

The relative entropy of coherence satisfies \cite{rastgro17}
\begin{align}
&h_{1}(P_{\suc})\leq{C}_{1}(\bro)+S_{1}(\bro)
\nonumber\\
&\leq
P_{\suc}\,\ln\!\left(\frac{M}{P_{\suc}}\right)+
(1-P_{\suc})\,\ln\!\left(\frac{N-M}{1-P_{\suc}}\right)
 , \label{renp02}
\end{align}
where $h_{1}(P_{\suc})$ is the binary Shannon entropy. Calculating
$C_{1}\bigl(\bro(t)\bigr)$, we examine it from the viewpoint of
(\ref{renp02}). To do so, we recall the complete form of
$\bro(t)$, namely
\begin{align}
\bro(t)=&\,\frac{1+r_{z}(t)}{2}\> |w\rangle\langle{w}|
+\frac{r_{x}(t)}{2}\,\bigl(|w\rangle\langle{m}|+|m\rangle\langle{w}|\bigr)
\nonumber\\
&+\frac{1-r_{z}(t)}{2}\> |m\rangle\langle{m}|
\, . \label{rhtwm}
\end{align}
In the computational basis, the diagonal part of $\bro(t)$ can be
written as the diagonal matrix, which has the value
$\bigl(1-P_{\suc}(t)\bigr)/(N-M)$ with multiplicity $N-M$ and the
value $P_{\suc}(t)/M$ with multiplicity $M$. For $t>0$, the
non-zero eigenvalues of $\bro(t)$ are obtained as
\begin{equation}
\frac{1\pm\|\vcr(t)\|_{2}}{2}
\ , \qquad
\|\vcr(t)\|_{2}=\sqrt{r_{x}(t)^{2}+r_{z}(t)^{2}}<1
\, . \label{eigrt}
\end{equation}
Of course, with the initial distribution (\ref{uniam}) we have
$\|\vcr(0)\|_{2}=1$. Due to (\ref{prpxz}), we have
$\|\vcr(t)\|_{2}\propto\eta^{t/2}$. In the considered case of
amplitude amplification, we have the equality
\begin{align}
C_{1}\bigl(\bro(t)\bigr)=
&\>P_{\suc}(t)\,\ln\!\left(\frac{M}{P_{\suc}(t)}\right)
\nonumber\\
&
+\bigl(1-P_{\suc}(t)\bigr)\ln\!\left(\frac{N-M}{1-P_{\suc}(t)}\right)-S_{1}\bigl(\bro(t)\bigr)
\, . \nonumber
\end{align}
That is, the upper bound of the right-hand side of (\ref{renp02})
is saturated here. For the given $P_{\suc}$, this upper bound
approves the maximal possible value of $C_{1}(\bro)$. We see that
trade-offs between $C_{1}\bigl(\bro(t)\bigr)$ and $P_{\suc}(t)$
follow the mentioned line. Let us exemplify a dependence of the
relative entropy of coherence on the step number $t$. In Fig.
\ref{fig3}, we take $N=64$, $M=1$, and show
$C_{1}\bigl(\bro(t)\bigr)$ for the four values of $\eta$. Except
for the value $\eta=1.0$, the curves asymptotically lie on the
constant line. This constant is generally written as
$(1/2)\ln\bigl(MN-M^{2}\bigr)$. Substituting $N=64$ and $M=1$, we
have the value $\ln63/2\approx2.072$, which is also seen in Fig.
\ref{fig3}. For the value $\eta=1$, we may compare the two curves
in Figs. \ref{fig1} and \ref{fig3}. One observes that peaks of
$P_{\suc}(t)$ correspond to valleys of $C_{1}\bigl(\bro(t)\bigr)$,
and {\it vice versa}. In detail, this property was discussed in
\cite{hfan2016}. We also observe that the curves for $\eta<1$
reveal a similar behavior but now with decaying. Even if the
amount of phase noise is low, oscillations of of the relative
entropy are reduced sufficiently quickly. These results
additionally maintain the conclusions previously reported in
\cite{rastgro17}. Namely, even tight trade-off relation between
coherence and the success probability does not imply a high
quality of amplitude amplification. This question deserves further
investigations.

\section{Conclusions}\label{sec6}

We have examined the case, when queries to the oracle in Grover's
search algorithm are exposed to phase distortions of the specific
type. Another possibility is that the oracle-box wires are altered
due to intrusion of an opposite party. The used model of
collective phase flips is similar to the phase damping channel.
Despite of its simplicity, this model allows us to observe some
genuine features of amplitude amplification processes. We have
concluded that Grover's search algorithm is actually sensitive to
collective phase flips occurring in the oracle-box wires. This
feature also provides an opposite party with chances to prevent
proper queries of legitimate users to the oracle. At the same
time, phase flips are such that the value of the success
probability is not changed during transfer via these wires. Even
if the user has been ensured with testing states, he is hardly
able to detect such distortions by means of one-time queries to
the oracle. We also investigated trade-off relation between
coherence and the success probability under noise of the
considered type. Our findings have further supported the conclusions 
obtained previously.

\appendix

\section{Solutions of the recursion equation}\label{app1}

To solve (\ref{itrr}), we calculate the eigenvalues and the
corresponding eigenvectors. We begin with the case (i), when
$\eta>A_{+}^{2}\,$. By calculations, one has
\begin{equation}
\cos2\theta=1-\frac{8M}{N}+\frac{8M^{2}}{N^{2}}
\ . \label{cos2t}
\end{equation}
The most interesting case occurs, when $M\ll{N}$ and the term
$\cos2\theta$ is sufficiently close to $1$. The condition of the
case (i),
\begin{equation}
\frac{2\,\sqrt{\eta}}{1+\eta}
>\cos2\theta
\, , \label{condi}
\end{equation}
will be fulfilled for $\eta_{\min}<\eta<1$ with
\begin{equation}
\sqrt{\eta_{\min}}=\frac{1-\sin2\theta}{\cos2\theta}
\ . \nonumber
\end{equation}
The value $\eta_{\min}$ does not approach $1$ with necessity,
whence the above range may be wide enough. Say, for $M=1$ and
$N=64$ we get $\cos2\theta\approx0.877$ and
$\eta_{\min}\approx0.351$. Due to $B^{2}=\eta-A_{+}^{2}$, the
eigenvalues are written in the form
\begin{equation}
\lambda_{\pm}=A_{+}\pm\iu{B}
\, . \label{eigvs0}
\end{equation}
Using $A_{+}^{2}+B^{2}=\eta$, put positive angle $\varphi$ such
that
\begin{align}
\frac{A_{+}}{\sqrt{\eta}}&=\cos\varphi
\, , \qquad
\frac{B}{\sqrt{\eta}}=\sin\varphi
\, , \label{phic0}\\
\varphi&=\arctan\!\left(\frac{B}{A_{+}}\right)
 . \label{phdf0}
\end{align}
The eigenvalues are rewritten as
$\lambda_{\pm}=\sqrt{\eta}\,\exp(\pm\iu\varphi)$. Calculating the
corresponding eigenvectors, we further obtain
\begin{equation}
\mtx^{-1}\mtm\mtx=\mtd
\, , \label{dagb}
\end{equation}
where $\mtd=\dig(\lambda_{+},\lambda_{-})$ and
\begin{align}
\mtx&=
\begin{pmatrix}
    \sin2\theta & \sin2\theta \\
    A_{-}+\iu{B} & A_{-}-\iu{B}
\end{pmatrix}
 , \label{xxp0}\\
\mtx^{-1}&=
\frac{1}{2\iu{B}\sin2\theta}
\begin{pmatrix}
    \iu{B}-A_{-} & \sin2\theta \\
    \iu{B}+A_{-} & -\sin2\theta
\end{pmatrix}
 . \label{xxm0}
\end{align}
Calculations of the matrix $\mtm^{t}=\mtx\mtd^{t}\mtx^{-1}$
finally give
\begin{equation}
\frac{\eta^{t/2}}{B}
\begin{pmatrix}
B\cos\varphi{t}-A_{-}\sin\varphi{t} & \sin\varphi{t}\,\sin2\theta \\
    -\,\eta\,\sin\varphi{t}\,\sin2\theta & B\cos\varphi{t}+A_{-}\sin\varphi{t}
\end{pmatrix}
 . \nonumber
\end{equation}
Due to $r_{x}(0)=\sin\theta$ and $r_{z}(0)=\cos\theta$, the above
formulas result in (\ref{arzt0}) and (\ref{psct0}). For $\eta=1$,
when the oracle-we have $A_{+}=\cos2\theta$, $A_{-}=0$,
$B=\sin2\theta$, and $\varphi=2\theta$. Then the expression
(\ref{psct0}) is reduced to
\begin{equation}
P_{\suc}(t)=\frac{1-\cos(2\theta{t}+\theta)}{2}
=\sin^{2}\bigl[\theta(t+1/2)\bigr]
\, . \label{psct00}
\end{equation}
The latter is well known for the original Grover algorithm. It is
seen from the formula for $\mtm^{t}$ that components of the Bloch
vector are proportional to the factor $\eta^{t/2}$, namely
\begin{equation}
r_{x}(t)\propto\eta^{t/2}
\, ,  \qquad
r_{z}(t)\propto\eta^{t/2}
\, .  \label{prpxz}
\end{equation}
Except for the value $\eta=1$, these components asymptotically
tends to zero. Hence, the probability $P_{\suc}(t)$ goes to $1/2$.

Let us proceed to the case (ii), when $\eta<A_{+}^{2}\,$. Due to
$B^{2}=A_{+}^{2}-\eta$, the eigenvalues are expressed as
\begin{equation}
\lambda_{\pm}=A_{+}\pm{B}
\, . \label{eigvs}
\end{equation}
For $\eta>0$, the eigenvalues are both strictly positive and
$\lambda_{-}<\lambda_{+}\leq{A}_{+}+A_{-}=\cos2\theta<1$. So, the
matrix $\mtm$ describes a contracting map. Due to
$A_{+}^{2}-B^{2}=\eta$, we can write
\begin{equation}
\frac{A_{+}}{\sqrt{\eta}}=\cosh\phi
\, , \qquad
\frac{B}{\sqrt{\eta}}=\sinh\phi
\, , \label{phic1}
\end{equation}
where positive parameter $\phi$ reads as
\begin{equation}
\phi=\frac{1}{2}\> \ln\!\left(\frac{A_{+}+B}{A_{+}-B}\right)
 . \label{phdf}
\end{equation}
Hence, we can write $\lambda_{\pm}=\sqrt{\eta}\,\exp(\pm\phi)$.
Similarly to the case (i), we diagonalize $\mtm$ according to
(\ref{dagb}). Now, the matrix of column eigenvectors and its
inverse are represented as
\begin{align}
\mtx&=
\begin{pmatrix}
    \sin2\theta & \sin2\theta \\
    A_{-}+B & A_{-}-B
\end{pmatrix}
 , \label{xxp1}\\
\mtx^{-1}&=
\frac{1}{2B\sin2\theta}
\begin{pmatrix}
    B-A_{-} & \sin2\theta \\
    B+A_{-} & -\sin2\theta
\end{pmatrix}
 . \label{xxm1}
\end{align}
Calculating the matrix $\mtm^{t}=\mtx\mtd^{t}\mtx^{-1}$, we finally
have
\begin{equation}
\frac{\eta^{t/2}}{B}
\begin{pmatrix}
B\cosh\phi{t}-A_{-}\sinh\phi{t} & \sinh\phi{t}\,\sin2\theta \\
    -\,\eta\,\sinh\phi{t}\,\sin2\theta & B\cosh\phi{t}+A_{-}\sinh\phi{t}
\end{pmatrix}
 , \nonumber
\end{equation}
since $A_{-}^{2}-B^{2}=\eta\sin^{2}2\theta$. Due to
$r_{x}(0)=\sin\theta$ and $r_{z}(0)=\cos\theta$, we finally get
(\ref{arzt}) and (\ref{psct}). The original Grover search is
beyond the case (ii). Indeed, for $1\leq{M}\leq{N}/2$ we have
$\cos2\theta<1$, so that the condition
$2\,\sqrt{\eta}<(1+\eta)\cos2\theta$ is certainly violated with
$\eta=1$.

\end{document}